\documentclass[preprintnumbers]{revtex4}

\usepackage{graphicx}
\def\ie{{\it i.e.}}

\def\mpl{\ifmmode \overline M_{Pl}\else $\overline M_{Pl}$\fi}
\def\to{\rightarrow}

\begin{document}
\bibliographystyle{revtex}

\preprint{SLAC-PUB-8959/
          P3-01}

\title{Extra Dimensional Signatures at CLIC}

\author{Thomas G. Rizzo}

\email[]{rizzo@slac.stanford.edu}
\affiliation{Stanford Linear Accelerator Center, 
Stanford University, Stanford, California 94309 USA}

\date{\today}

\begin{abstract}
A brief overview is presented of the signatures for several different models 
with extra dimensions at CLIC, an $e^+e^-$ linear collider with a center of 
mass energy of 3-5 TeV and an integrated luminosity of order 1~$ab^{-1}$. In 
all cases the search reach for the resulting new physics signatures is found 
to be in the range of $\simeq$15-80 TeV.
\end{abstract}

\maketitle

\section{Introduction}

Many models predict the existence of additional spatial dimensions that 
lead to new and distinct phenomenological signatures for future colliders which 
have center of mass energies in the TeV range and above. Most of the models 
in the literature fall into one of the three following classes: 
($i$) the large extra dimensions scenario of 
Arkani-Hamed, Dvali and Dimopoulos(ADD){\cite {add}}. This model predicts the 
emission and exchange of large Kaluza-Klein(KK) towers of gravitons that are 
finely-spaced in mass. The emitted gravitons appear as missing energy while 
the KK tower exchange leads to contact interaction-like dimension-8 operators. 
($ii$) A second possibility are models where the extra dimensions are of TeV 
scale in size. In these scenarios there are KK excitations of the SM gauge 
(and possibly other SM) fields with masses of order a TeV which can appear as 
resonances at colliders. ($iii$) A last class of models are those with warped 
extra dimensions, such as the Randall-Sundrum Model(RS){\cite {rs}}, which 
predict graviton resonances with both weak scale masses and couplings to 
matter. High energy lepton colliders in the multi-TeV range with sufficient 
luminosity, such as CLIC, will be able to both directly and indirectly search 
for and/or make detailed studies of models in all three classes. The case 
of direct searches is rather straightforward as we are producing the 
new physics, such as a KK resonance, directly. Indirect searches 
are more subtle 
but the capability of making high precision measurements at lepton colliders 
allows us to probe mass scales far in excess of the collider center of mass 
energy, in some cases by more than an order of magnitude. For most models of 
type ($i$) or ($iii$) which deal with the hierarchy problem, 
if no signal is observed by the time the mass 
scales probed by CLIC are reached, the motivation behind these particular 
models will be greatly weakened if not entirely removed. 
In what follows, for simplicity, we will only focus on searches involving 
the process $e^+e^- \to f\bar f$. From studies performed for both  
NLC/TESLA and LEP we know that this channel provides an excellent probe of 
the parameter spaces of extra-dimensional models and we expect that this will 
continue to be true at even higher energies.

\section{Signatures}

The first model we consider is ADD; we will limit our discussion to 
the case of graviton tower exchange in $e^+e^-\to f\bar f$. 
The effect of summing the KK gravitons is to produce a set of effective 
dimension-8 operators of the form $\sim \lambda T^{\mu\nu}T_{\mu\nu}/M_s^4$, 
where $T_{\mu\nu}$ is the stress-energy tensor of the SM matter exchanging 
the tower{\cite {pheno}}. 
This approximation only applies in the limit that the center of 
mass energy of the collision process lies sufficiently below the cut-off 
scale, $M_s$, which is of order the size of the Planck scale in the extra 
dimensional space. In the convention used by Hewett{\cite {pheno}} and 
adopted here, the contribution of the spin-2 exchanges can be universally 
expressed in terms of the scale, $M_s$, and a sign, $\lambda$.
Current experimental constraints 
from LEP and the Tevatron{\cite {greg}} tell us that $M_s \geq 1$ TeV for 
either sign of $\lambda$; values for $M_s$ as 
large as the low 10's of TeV may be contemplated in this scenario.

\begin{figure}[htbp]
\centerline{
\includegraphics[width=5.4cm,angle=90]{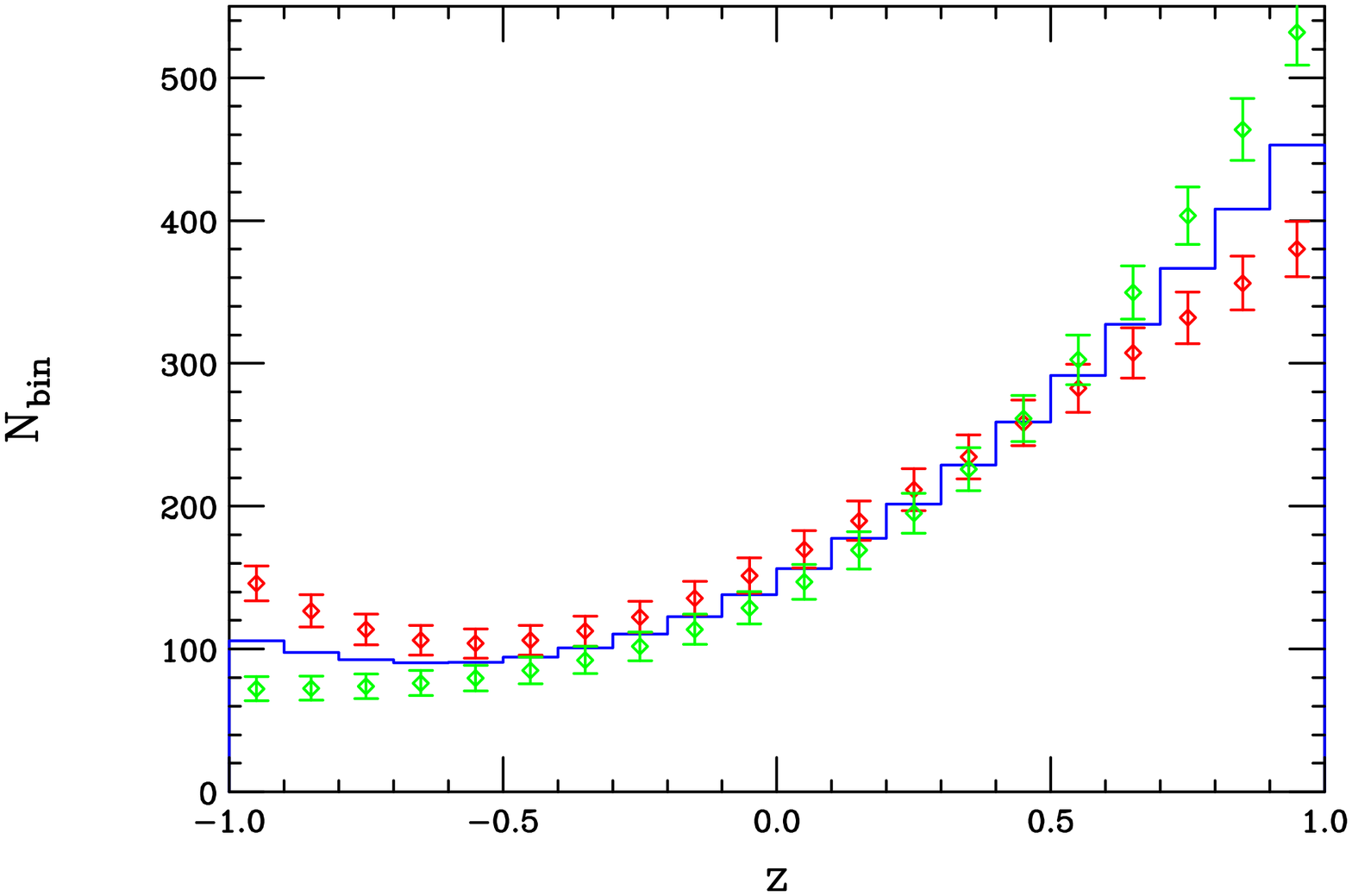}
\hspace*{5mm}
\includegraphics[width=5.4cm,angle=90]{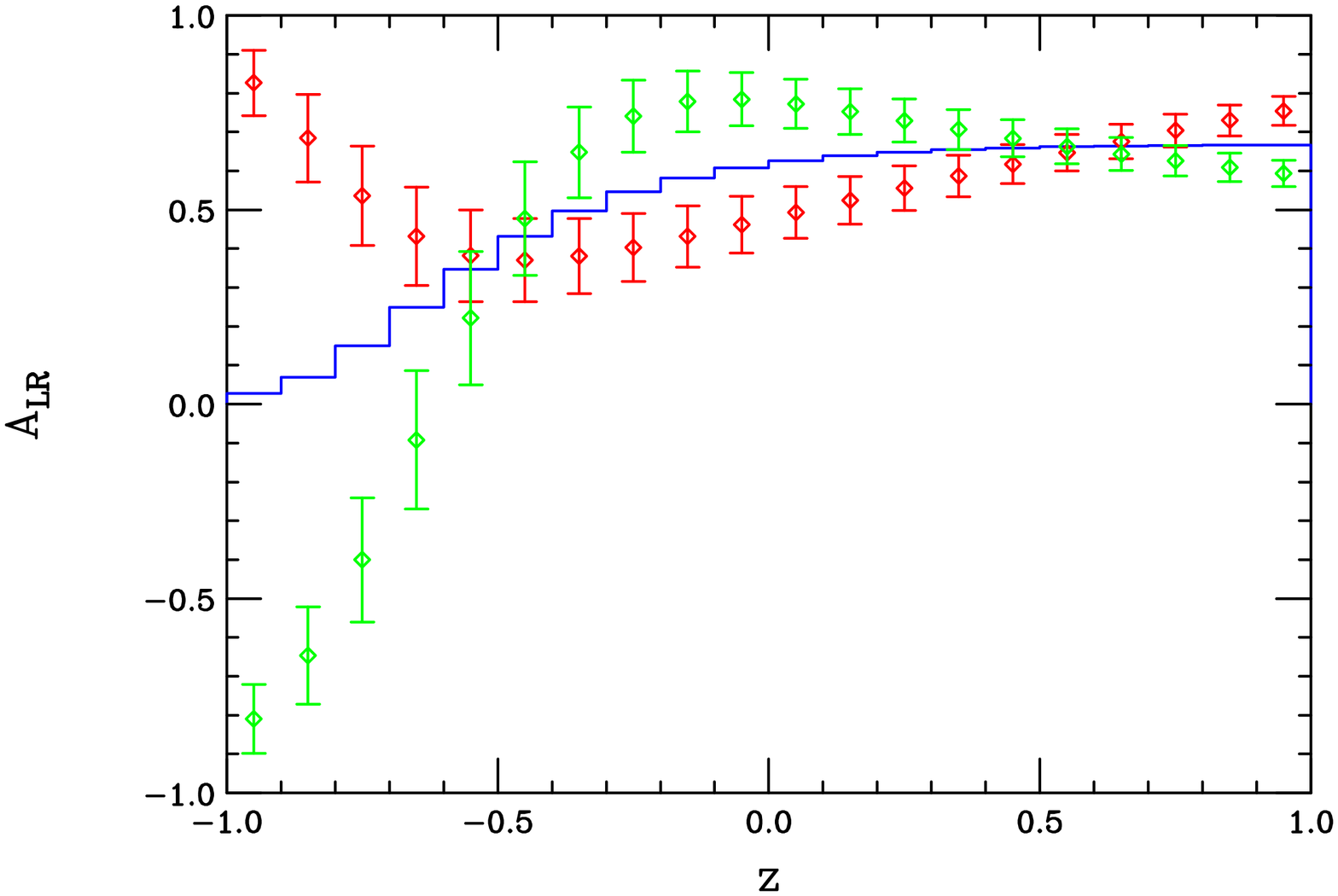}}
\vspace*{0.1cm}
\caption{Deviations in the cross section for $\mu$-pairs(left) and $A_{LR}$ 
for $b$-quarks(right) at $\sqrt s$=5 TeV for $M_s=15$ TeV in the ADD model 
for an integrated 
luminosity of 1$~ab^{-1}$. The SM is represented by the histogram while the 
red and green data points show the ADD predictions with $\lambda=\pm 1$. In 
both plots $z=\cos \theta$.}
\label{p3-01_fig1}
\end{figure}

\begin{figure}[htbp]
\centerline{
\includegraphics[width=5.4cm,angle=90]{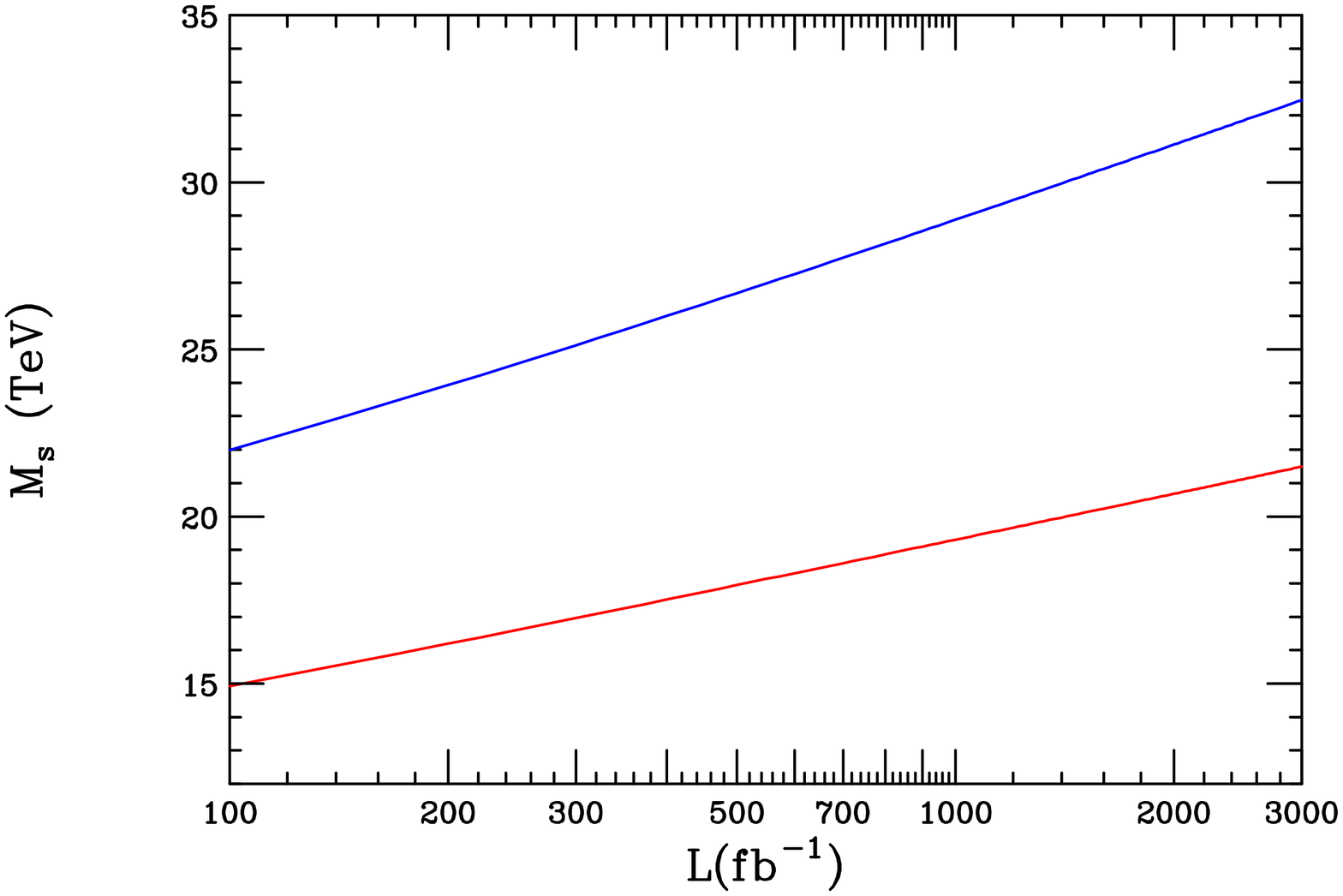}
\hspace*{5mm}
\includegraphics[width=5.4cm,angle=90]{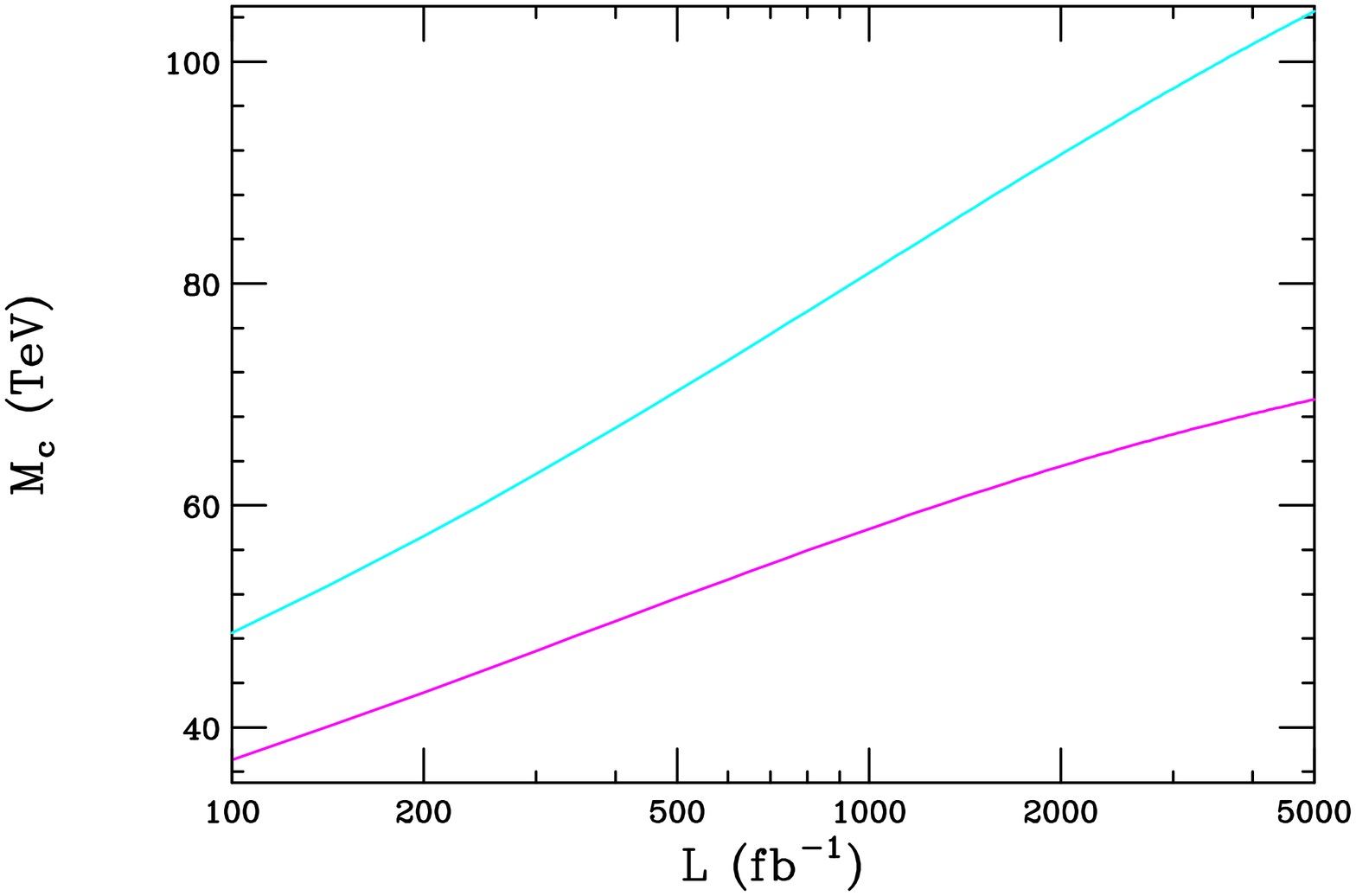}}
\vspace*{0.1cm}
\caption{(Left) Search reach for the ADD model scale $M_s$ at CLIC as a 
function of the integrated luminosity from the set of processes $e^+e^-\to 
f\bar f$ assuming $\sqrt s=3$(red) or 5(blue) TeV. Here $f=\mu,\tau,b,c$,  
$t$, etc. (Right) Corresponding reach for the compactification scale of the KK 
gauge bosons in the case of one extra dimension and all fermions localized 
at the same orbifold fixed point.
}
\label{p3-01_fig2}
\end{figure}

In the case of $e^+e^-\to f\bar f$, the addition of KK tower exchange leads to 
significant deviations in differential cross sections and polarization 
asymmetries from their SM values which are strongly dependent 
on both the sign of $\lambda$ and the ratio $s/M_s^2$. 
Such shifts are observable in final states of all flavors. In addition, the 
shape of these deviations from the SM with varying energy and 
scattering angle, as shown by Hewett{\cite {pheno}}, tells us that the 
underlying physics arises due to dimension-8 operators and not, for example, 
$Z'$ exchange. 
Fig.~\ref{p3-01_fig1} shows an example of how such deviations from the SM 
might appear at a 5 TeV CLIC in the case that $M_s$=15 TeV for either sign of 
$\lambda$. 
The indirect search reach for the scale $M_s$ can be obtained by combining the 
data for several of the fermion final states($\mu, \tau, c,b,t$, etc) in a 
single overall fit. The result of this analysis for CLIC is the $\lambda$ 
independent bound shown in 
Fig.~\ref{p3-01_fig2} as a function of the integrated luminosity for 
$\sqrt s$=3 or 5 TeV. For an integrated luminosity of 1 $ab^{-1}$ we see that 
the reach is $M_s\simeq 6\sqrt s$ which is consistent with analyses at lower 
energy machines{\cite {pheno}}. 

\begin{figure}[htbp]
\centerline{
\includegraphics[width=5.4cm,angle=90]{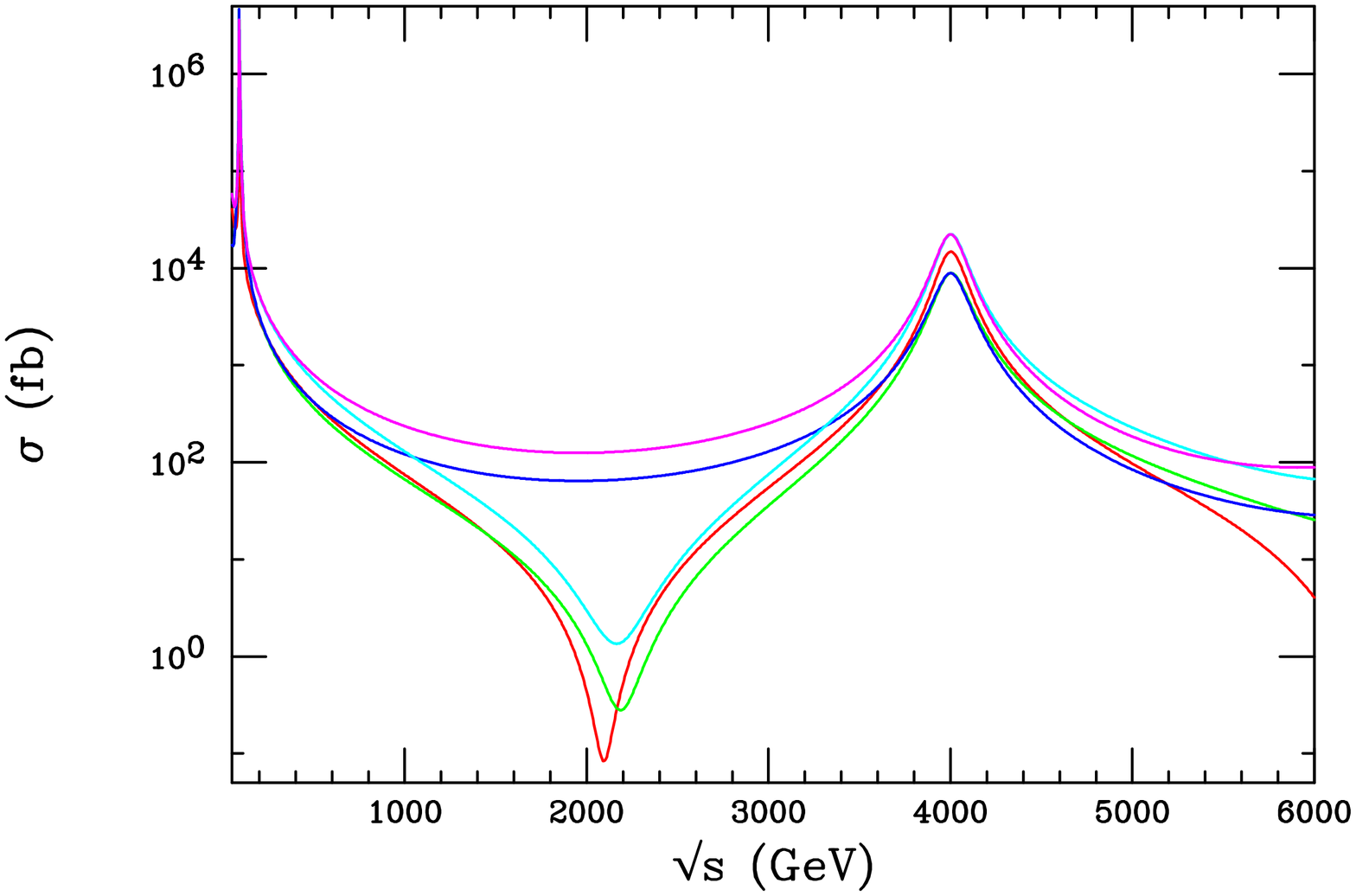}
\hspace*{5mm}
\includegraphics[width=5.4cm,angle=90]{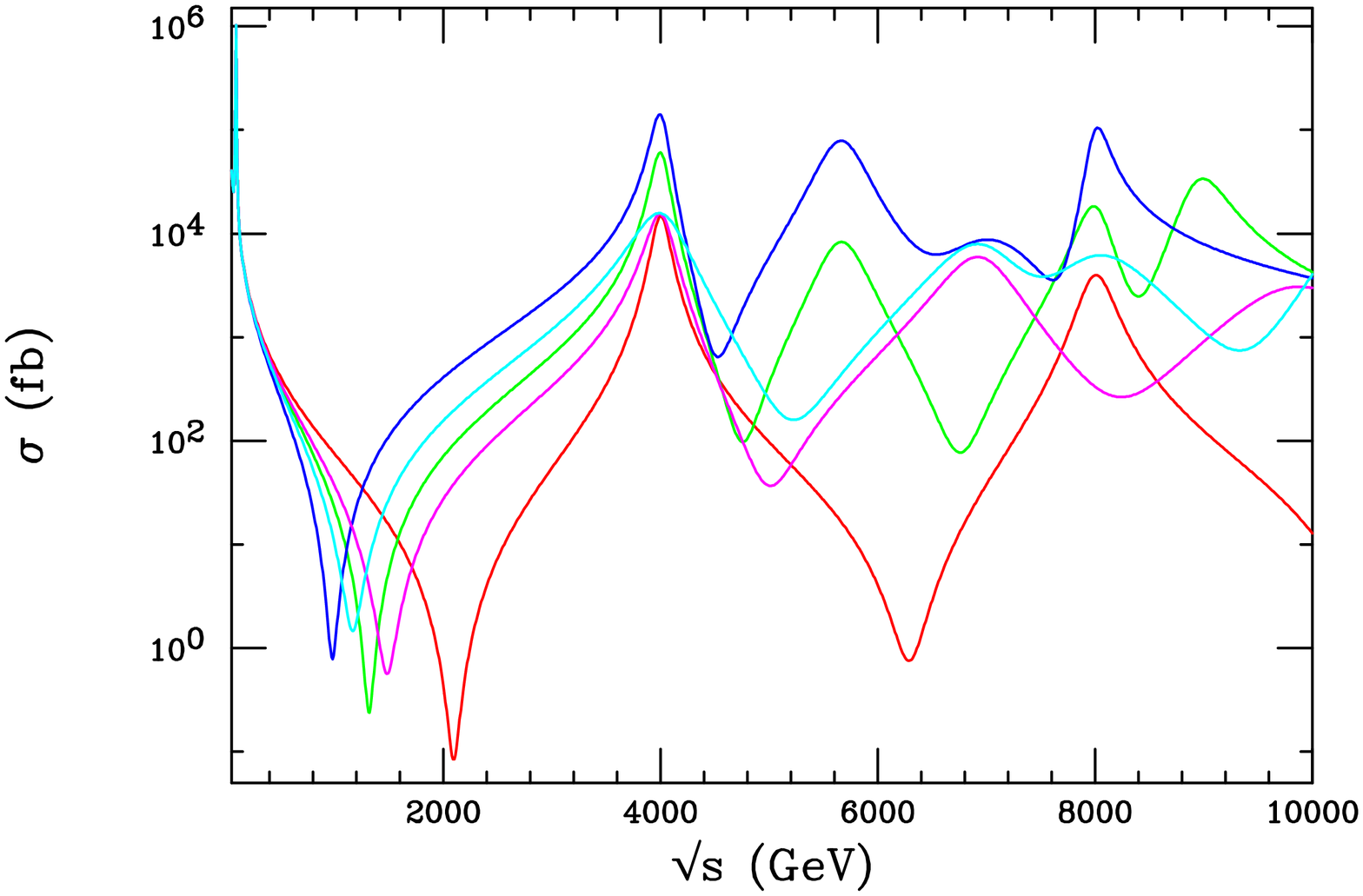}}
\vspace*{0.1cm}
\caption{(Left) Comparison of $e^+e^-\to f\bar f$ cross sections in the case 
of one extra dimension when $M_c=4$ TeV. The red curve is for the case $f=\mu$ 
while the green(blue) and cyan(magenta) curves are for the cases $f=b,c$, 
respectively when $D=0(\pi R_c)$. (Right) $e^+e^-\to \mu^+\mu^-$ cross sections 
for several different models with one or more extra dimensions assuming 
$M_c=4$ TeV.}
\label{p3-01_fig3}
\end{figure}

Next we turn to models with TeV scale extra dimensions. 
In the simplest versions of these theories, only the 
SM gauge fields are in the bulk whereas the fermions remain at one of 
the two orbifold 
fixed points; Higgs fields may lie at the fixed points or propagate in the 
bulk. (More complicated scenarios with very different phenomenology 
are possible.) It is possible that, \ie, quarks and leptons may lie at 
{\it different} fixed points in which case they would be separated by a 
distance $D=\pi R_c$, where $R_c$ is the compactification radius. 
In the case with only one extra dimension it has been shown that the 
current high precision 
electroweak data can place a lower bound on the mass of the first KK excited 
gauge boson in excess 
of $\simeq$ 4 TeV{\cite {bunch}}. In such a model, to a good 
approximation, the masses of the KK tower states are given by $M_n=nM_c$, 
where $M_c=R_c^{-1}$ is the compactification scale. 
For this one extra dimensional example all of the excited KK states 
have identical couplings to the SM fermions, apart from possible overall 
signs. In this case, only the first 
KK state may be observable at the LHC since KK modes with masses in excess of 
$\simeq 7$ TeV will be too massive to be produced. 
High energy $e^+e^-$ colliders can search for SM gauge boson excitations in 
exactly the same way as described above for ADD graviton tower exchange but 
with a significantly higher search reach, as shown in Fig.~\ref{p3-01_fig2}, 
since the shifts in SM observables are now due to effective dimension-6 
(instead of dimension-8) operators. Note that the search reach in this case 
can be as large as $\sim 15\sqrt s$. 
A very high energy CLIC may be even more useful if the number of extra 
dimensions is greater than one; in this case, still keeping the fermions at 
the orbifold fixed points, the bounds from precision data are expected to be 
stricter than in the one-dimensional case but are less quantitatively 
precise since the naive evaluation of the relevant 
sums over KK states are divergent.  
One now finds that the masses and couplings of 
KK excitations become both level and compactification-scheme dependent thus 
leading to a rather complex KK spectrum. Some sample KK excitation spectra for 
several different TeV-scale models with more than one extra 
dimension are shown in Fig.~\ref{p3-01_fig3}. Note that the 
measurements of the locations of the peaks and their relative heights and 
widths can be used to uniquely identify a given extra-dimensional model.

\begin{figure}[htbp]
\centerline{
\includegraphics[width=5.4cm,angle=90]{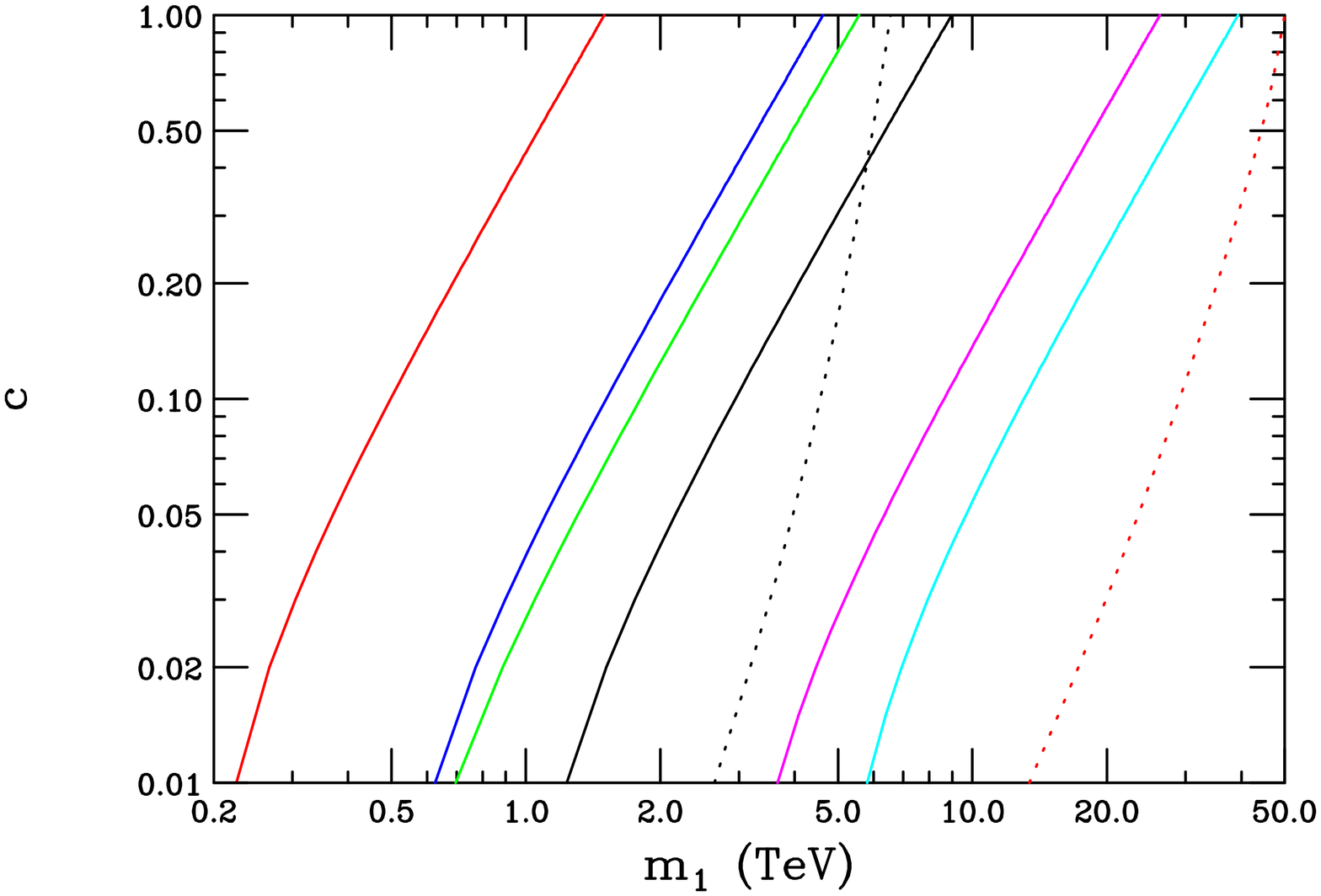}
\hspace*{5mm}
\includegraphics[width=5.4cm,angle=90]{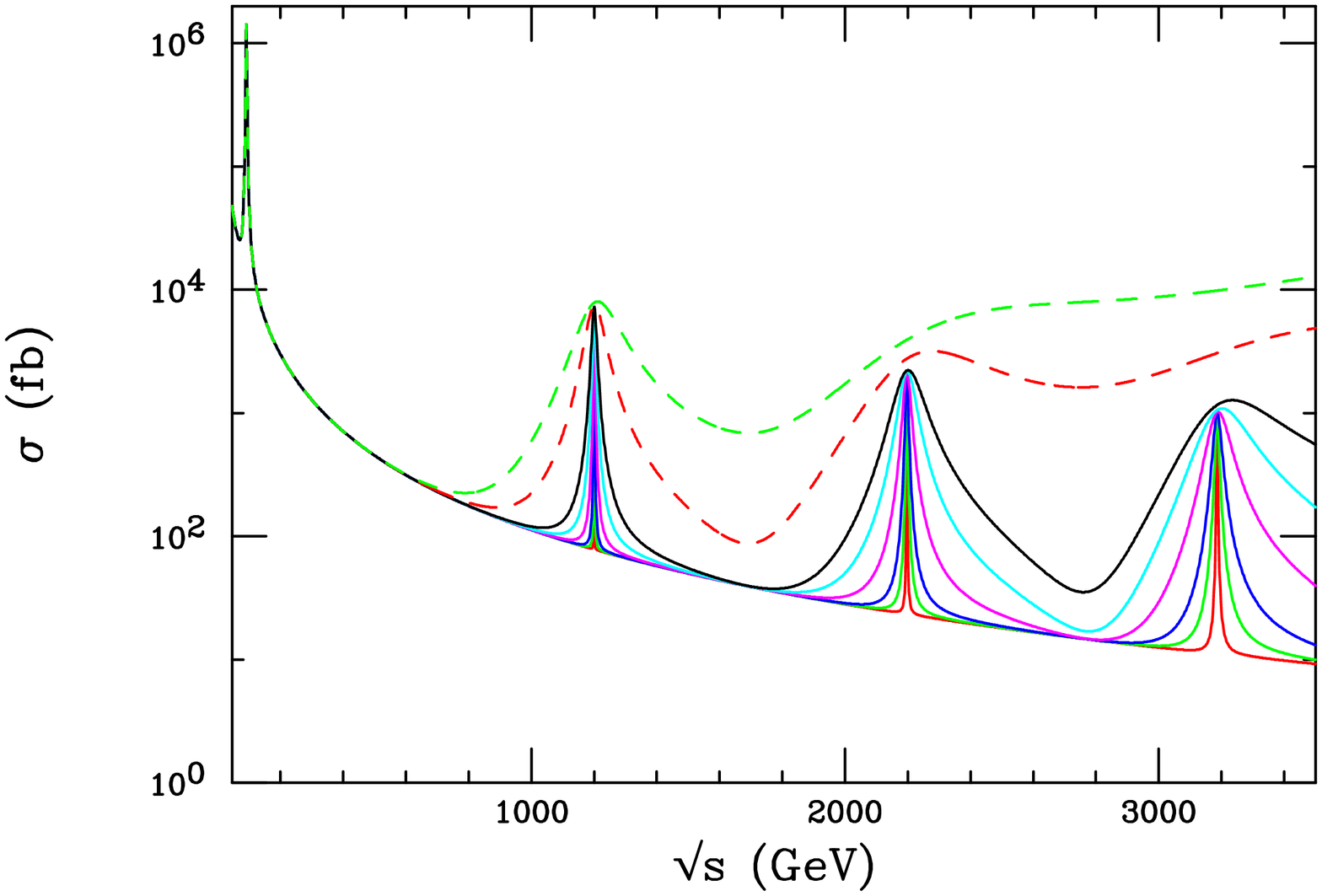}}
\vspace*{0.1cm}
\caption{(Left) Indirect constraints from $e^+e^-$ colliders on the RS model 
parameter space with $c=k/\mpl$; the excluded region is to the left of the 
curves. From left to right the solid curves correspond to bounds from LEP II, 
a 500 GeV LC with 75 or 500 $fb^{-1}$ luminosity, a TeV machine with 200 
$fb^{-1}$, and a 3 or 5 TeV CLIC with 1 $ab^{-1}$. The dotted lines are the 
corresponding LHC (100 $fb^{-1}$) and $\sqrt s=175$ TeV VLHC(200 $fb^{-}$) 
direct search reaches. (Right) KK graviton excitations in the RS model 
produced in the process $e^+e^-\to \mu^+\mu^-$. From the most narrow to widest 
resonances the curves are for $c$ in the range 0.01 to 0.2.}
\label{p3-01_fig4}
\end{figure}

\begin{figure}[htbp]
\centerline{
\includegraphics[width=5.4cm,angle=90]{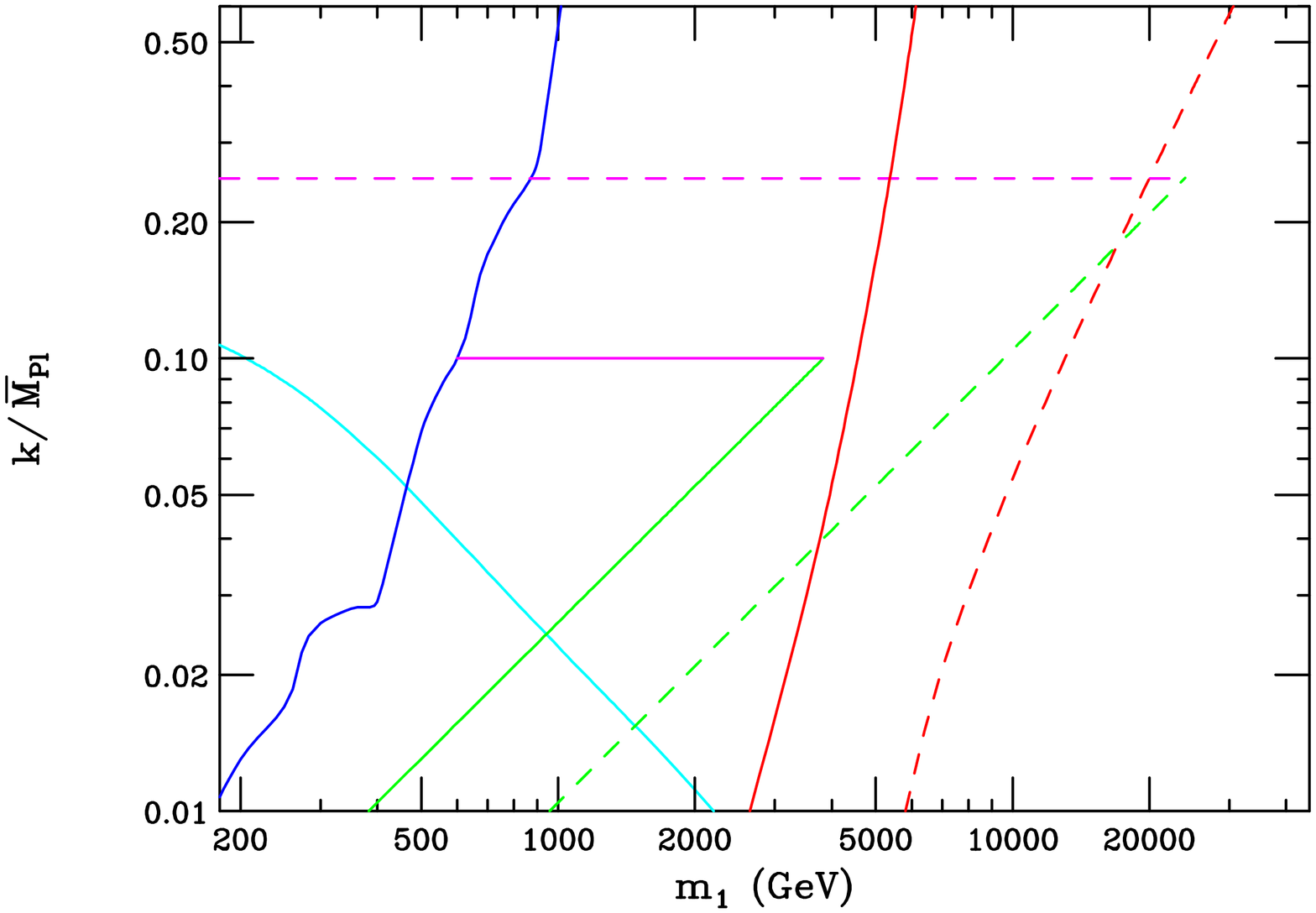}
\hspace*{5mm}
\includegraphics[width=5.4cm,angle=90]{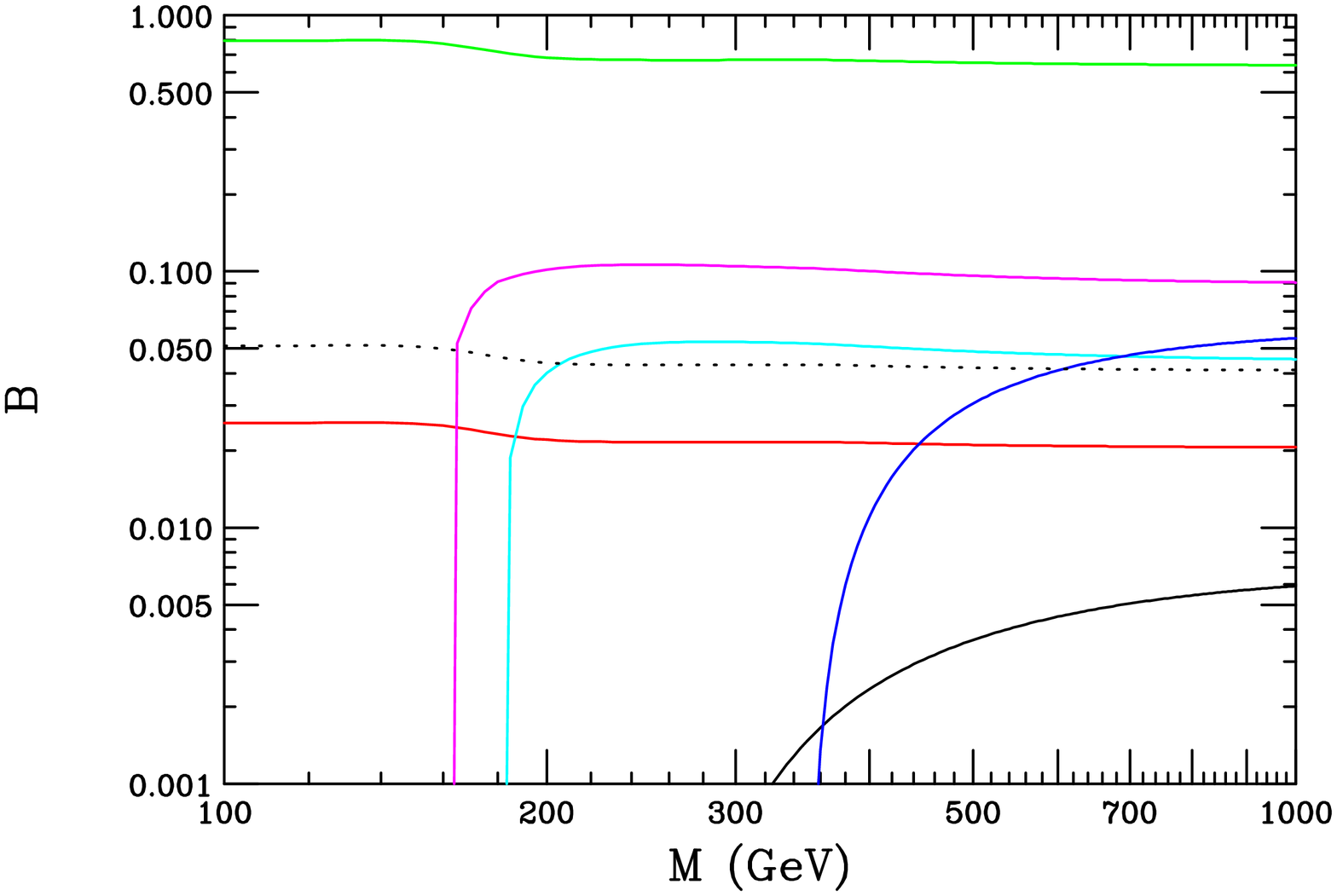}}
\vspace*{0.1cm}
\caption{(Left) Allowed regions of the RS model parameter space. Current 
Tevatron(blue) and precision measurements(cyan) forbid regions to the left 
of their specific curves. The horizontal magenta solid(dashed) lines form the 
upper bound of the region when $c=0.10(0.25)$ while the solid(dashed) green 
curve is the corresponding lower bound when $\Lambda_\pi=10(25)$ TeV. The 
solid(dashed) red curve is the reach of the LHC(CLIC) with 100 $fb^{-1}$
($\sqrt s=5 $ TeV with 1 $ab^{-1}$). (Right) Branching fractions for the 
lightest RS KK graviton; from top to bottom on the right-hand side the curves 
are for 2 jets, $W^+W^-$, $t\bar t$, $2Z$, $2\gamma$, $e^+e^-$ and $2h$, 
respectively.}
\label{p3-01_fig5}
\end{figure}

The last case we consider is the RS model wherein, as discussed above, we 
expect to produce TeV-scale graviton resonances in many channels{\cite {dhr}} 
including $e^+e^-\to f\bar f$. In its simplest version, with two branes, one 
extra dimension, 
and with all of the SM fields remaining on the TeV-brane, this model has only 
two fundamental parameters: the mass of the first KK state (from which all 
the others can be determined) and an additional 
parameter, $c=k/\mpl$, which we expect to be smaller than but not too far 
from unity. This parameter essentially controls the effective coupling 
strength of the gravitons(when expressed in terms of the mass of the lowest 
lying KK state) and thus also the widths of the corresponding resonances. 
Below the mass of the lightest resonance linear colliders can still search 
indirectly for the contributions of RS graviton exchange in a manner similar 
to that described above; the results of such an analysis are shown in 
Fig.~\ref{p3-01_fig4}. On top of the resonances as in Fig.~\ref{p3-01_fig4} 
the decay angular distribution can be easily determined allowing us to 
demonstrate that a spin-2 particle is being produced while measurements of 
the branching fractions to various decay modes, shown in Fig.~\ref{p3-01_fig5}, 
would prove that we 
are producing gravitons. If several resonances are produced the ratios of 
their masses can be used verify the RS scenario since their masses are in the 
ratios of the roots of the $J_1$ Bessel function. 
It also seems likely that CLIC will be able to 
perform a detailed study of some of the more exotic decays of the heavier 
graviton states{\cite {us}} that may occur in this model. 
Fig.~\ref{p3-01_fig5} shows the current bounds on the RS parameter space from 
both precision measurements and Tevatron searches. Also shown are the 
constraints from naturalness on $\Lambda_\pi$ and on $c$ from the requirement 
of stability under quantum corrections. For the tighter set of constants the 
LHC can cover all of the model space whereas if these theoretical constraints 
are allowed to be somewhat weakened then the whole space will be essentially 
covered by CLIC.

\section{Discussion and Conclusion}

From the discussion above it is clear that the high center of mass energy of 
CLIC offers a great opportunity to study many different models with 
extra dimensions.

%
\def\MPL #1 #2 #3 {Mod. Phys. Lett. {\bf#1},\ #2 (#3)}
\def\NPB #1 #2 #3 {Nucl. Phys. {\bf#1},\ #2 (#3)}
\def\PLB #1 #2 #3 {Phys. Lett. {\bf#1},\ #2 (#3)}
\def\PR #1 #2 #3 {Phys. Rep. {\bf#1},\ #2 (#3)}
\def\PRD #1 #2 #3 {Phys. Rev. {\bf#1},\ #2 (#3)}
\def\PRL #1 #2 #3 {Phys. Rev. Lett. {\bf#1},\ #2 (#3)}
\def\RMP #1 #2 #3 {Rev. Mod. Phys. {\bf#1},\ #2 (#3)}
\def\NIM #1 #2 #3 {Nuc. Inst. Meth. {\bf#1},\ #2 (#3)}
\def\ZPC #1 #2 #3 {Z. Phys. {\bf#1},\ #2 (#3)}
\def\EJPC #1 #2 #3 {E. Phys. J. {\bf#1},\ #2 (#3)}
\def\IJMP #1 #2 #3 {Int. J. Mod. Phys. {\bf#1},\ #2 (#3)}
\def\JHEP #1 #2 #3 {J. High En. Phys. {\bf#1},\ #2 (#3)}

\end{document}